\documentclass{elsart} 
\journal{Physics Letters B}

\begin{document}
\begin{frontmatter}

\title{The cosmic jerk parameter in $f(R)$ gravity} 
\author[IU]{Nikodem J. Pop\l awski\corauthref{cor}}
\corauth[cor]{Corresponding author}
\ead{nipoplaw@indiana.edu}
\address[IU]{Department of Physics, Indiana University, 727 East Third Street, Bloomington, IN 47405, USA}

\begin{abstract}
We derive the expression for the jerk parameter in $f(R)$ gravity. 
We use the Palatini variational principle and the field equations in the
Einstein conformal gauge.
For the particular case $f(R)=R-\frac{\alpha^2}{3R}$, the predicted value of
the jerk parameter agrees with the SNLS SNIa and X-ray galaxy cluster
distance data but does not with the SNIa gold sample data.
\end{abstract}

\end{frontmatter}

\section{Introduction}

A particular class of alternative theories of gravity that has recently 
attracted a lot of interest is that of the $f(R)$ gravity 
models, in which the gravitational Lagrangian is a function of 
the curvature scalar $R$~\cite{fR}
It has been shown that current cosmic acceleration may originate from the
addition of a term $R^{-1}$ to the Einstein--Hilbert Lagrangian
$R$~\cite{acc}.

As in general relativity, $f(R)$ gravity theories obtain the field equations
by varying the total action for both the field and matter. 
In this work we use the metric--affine (Palatini) variational principle,
according to which the metric and connection are considered as 
geometrically independent quantities, and the action is varied 
with respect to both of them~\cite{SL}.
The other one is the metric (Einstein--Hilbert) variational principle, 
according to which the action is varied with respect to the metric 
tensor $g_{\mu\nu}$, and the affine connection coefficients are
the Christoffel symbols of $g_{\mu\nu}$.
Both approaches give the same result 
only if we use the standard Einstein--Hilbert action~\cite{Schr}.
The field equations in the Palatini formalism are second-order 
differential equations, while for metric theories they are 
fourth-order.
Another remarkable property of the metric--affine approach is that the 
field equations in vacuum reduce to the standard Einstein equations of 
general relativity with a cosmological constant~\cite{Schr}.

One can show that $f(R)$ theories of gravitation 
are conformally equivalent to the Einstein theory
of the gravitational field interacting with additional matter 
fields, if the action for matter does not depend on connection~\cite{SL,eq}. 
This can be done by means of a Legendre 
transformation, which in classical mechanics replaces 
the Lagrangian of a mechanical system with the Helmholtz 
Lagrangian. 
For $f(R)$ gravity, the scalar degree of freedom 
due to the occurrence of nonlinear second-order terms in the Lagrangian 
is transformed into an auxiliary scalar field $\phi$~\cite{eq}.
The set of variables $(g_{\mu\nu},\,\phi)$ is commonly called the 
{\it Jordan conformal gauge}. 
In the Jordan gauge, the connection is metric incompatible unless $f(R)=R$.
The compatibility can be restored by a certain conformal 
transformation of the metric: $g_{\mu\nu}\rightarrow 
h_{\mu\nu}=f'(R)g_{\mu\nu}$. 
The new set $(h_{\mu\nu},\,\phi)$ is called the {\it Einstein conformal 
gauge}, and we will regard the metric in this gauge as physical. 

$f(R)$ gravity models have been compared with cosmological observations by
several authors~\cite{obs,Niko2} and the problem of viability of
these models is still open (see~\cite{Far} and references therein).
Current SNIa observations provide the data on the time evolution of the
deceleration parameter $q$ in the form of $q=q(z)$, where $z$ is the
redshift~\cite{Gold}.
The extraction of the information from these data depends, however, on
assumed parametrization of $q(z)$~\cite{jerk2}.
For small values of $z$ such a dependence can be linear,
$q(z)=q_0+q_1 z$~\cite{Gold}, but its validity should fail at $z\sim 1$.
A convenient method to describe models close to $\Lambda CDM$ is based on
the cosmic jerk parameter $j$, a dimensionless third derivative of the scale
factor with respect to the cosmic time~\cite{jerk1,snap}.
A deceleration-to-acceleration transition occurs for models with
a positive value of $j_0$ and negative $q_0$.
Flat $\Lambda CDM$ models have a constant jerk $j=1$.

In this work we derive the general expression for the jerk parameter
in $f(R)$ gravity.
We use the field equations in the Palatini formalism and the Einstein
conformal gauge~\cite{Niko1}.
We find the current value of this parameter for the case
$f(R)=R-\frac{\alpha^2}{3R}$~\cite{acc,Niko2} and compare
it with recent cosmological data~\cite{jerk2}.

\section{Palatini variation in $f(R)$ gravity}

The action for $f(R)$ gravity in the original (Jordan) gauge 
with the metric $\tilde{g}_{\mu\nu}$ is given by~\cite{Niko1}
\begin{equation}
S_J=-\frac{1}{2\kappa c}\int d^4 
x\bigl[\sqrt{-\tilde{g}}f(\tilde{R})\bigr] + 
S_m(\tilde{g}_{\mu\nu},\psi).
\label{action1}
\end{equation}
Here, $\sqrt{-\tilde{g}}f(\tilde{R})$ is a Lagrangian density that depends 
on the curvature scalar 
$\tilde{R}=R_{\mu\nu}(\Gamma^{\,\,\lambda}_{\rho\,\sigma})\tilde{g}^{\mu\nu}$, 
$S_m$ is the action for matter represented 
symbolically by $\psi$ and independent of the connection, 
and $\kappa=\frac{8\pi G}{c^4}$. 
Tildes indicate quantities calculated in the Jordan gauge.

Variation of the action $S_J$ with respect to $\tilde{g}_{\mu\nu}$ 
yields the field equations
\begin{equation}
f'(\tilde{R})R_{\mu\nu}-\frac{1}{2}f(\tilde{R})\tilde{g}_{\mu\nu}=\kappa 
T_{\mu\nu},
\label{field1}
\end{equation} 
where the dynamical energy--momentum tensor of matter is generated 
by the Jordan metric tensor:
\begin{equation}
\delta S_m=\frac{1}{2c}\int d^4 x\sqrt{-\tilde{g}}\,T_{\mu\nu}\delta\tilde{g}^{\mu\nu},
\label{EMT1}
\end{equation} 
and the prime denotes the derivative of a function with respect to its variable.
From variation of $S_J$ with 
respect to the connection $\Gamma^{\,\,\rho}_{\mu\,\nu}$
it follows that this connection is given by the Christoffel 
symbols of the conformally transformed metric~\cite{eq} 
\begin{equation}
g_{\mu\nu}=f'(\tilde{R})\tilde{g}_{\mu\nu}.
\label{conf}
\end{equation}
The metric $g_{\mu\nu}$ defines the Einstein gauge, in which the connection is metric compatible.

The action~(\ref{action1}) is dynamically equivalent 
to the following Helmholtz action \cite{eq,Niko1}:
\begin{equation}
S_H=-\frac{1}{2\kappa c}\int d^4 x\sqrt{-\tilde{g}}\bigl[f(\phi(p))+p(\tilde{R}-\phi(p))\bigr]+S_m(\tilde{g}_{\mu\nu},\psi),
\label{action2}
\end{equation}
where $p$ is a new scalar variable.
The function $\phi(p)$ is determined by
\begin{equation}
\frac{\partial 
f(\tilde{R})}{\partial\tilde{R}}\bigg{\vert}_{\tilde{R}=\phi(p)}=p.
\label{phi}
\end{equation}
From Eqs.~(\ref{conf}) and~(\ref{phi}) it follows that
\begin{equation}
\phi=Rf'(\phi),
\label{resc}
\end{equation}
where $R=R_{\mu\nu}(\Gamma^{\,\,\lambda}_{\rho\,\sigma})g^{\mu\nu}$ is
the Riemannian curvature scalar of the metric $g_{\mu\nu}$.

In the Einstein gauge, the action~(\ref{action2}) becomes the standard 
Einstein--Hilbert action of general relativity with an additional
scalar field:
\begin{equation}
S_E=-\frac{1}{2\kappa c}\int d^4 x\sqrt{-g}\bigl[R-p^{-1}\phi(p)+p^{-2}f(\phi(p))\bigr]+S_m(p^{-1}g_{\mu\nu},\psi).
\label{action3}
\end{equation}
Choosing $\phi$ (which is the curvature scalar in the Jordan gauge)
as the scalar variable leads to
\begin{equation}
S_E=-\frac{1}{2\kappa c}\int d^4 x\sqrt{-g}\bigl[R-2V(\phi)\bigr]+S_m([f'(\phi)]^{-1}g_{\mu\nu},\psi),
\label{action4}
\end{equation}
where $V(\phi)$ is the effective potential
\begin{equation}
V(\phi)=\frac{\phi f'(\phi)-f(\phi)}{2[f'(\phi)]^2}.
\label{pot}
\end{equation}

Variation of the action~(\ref{action4}) with respect to 
$g_{\mu\nu}$ yields the equations of the gravitational field in
the Einstein gauge~\cite{Niko1}: 
\begin{equation}
R_{\mu\nu}-\frac{1}{2}Rg_{\mu\nu}=\frac{\kappa 
T_{\mu\nu}}{f'(\phi)}-V(\phi)g_{\mu\nu},
\label{EOF1}
\end{equation}
while variation with respect to $\phi$ reproduces~(\ref{resc}).
Eqs.~(\ref{resc}) and~(\ref{EOF1}) give
\begin{equation}
\phi f'(\phi)-2f(\phi)=\kappa Tf'(\phi),
\label{struc2}
\end{equation}
from which we obtain $\phi=\phi(T)$.
Substituting $\phi$ into the field equations~(\ref{EOF1}) leads
to a relation between the Ricci tensor and the energy--momentum tensor.
Such a relation is in general nonlinear and depends on the form of the
function $f(R)$.

\section{The jerk parameter in $f(R)$ gravity}

The jerk parameter in cosmology is defined as~\cite{jerk1,snap}
\begin{equation}
j=\frac{\dot{\ddot{a}}}{aH^3},
\label{jer1}
\end{equation}
where $a$ is the cosmic scale factor, $H$ is the Hubble parameter, and the
dot denotes differentiation with respect to the cosmic time.
This parameter appears in the fourth term of a Taylor expansion of the
scale factor around $a_0$:
\begin{eqnarray}
& & \frac{a(t)}{a_0}=1+H_0(t-t_0)-\frac{1}{2}q_0H^2_0(t-t_0)^2+\frac{1}{6}j_0H^3_0(t-t_0)^3 \nonumber \\
& & +\,O[(t-t_0)^4],
\label{exp}
\end{eqnarray}
where the subscript $0$ denotes the present value.
We can rewrite Eq.~(\ref{jer1}) as
\begin{equation}
j=q+2q^2-\frac{\dot{q}}{H},
\label{jer2}
\end{equation}
where $q$ is the deceleration parameter.
For a flat $\Lambda CDM$ model $j=1$~\cite{jerk2}.\footnote[1]{This identity
can be easily verified from Eq.~(\ref{jer2}) for special cases where
the deceleration parameter is constant: 
$q=1/2$ (matter dominated universe) and $q=-1$ (de Sitter universe).}

From the gravitational field equations~(\ref{EOF1}) applied to a flat
Robertson--Walker universe with dust
we can derive the $\phi$-dependence of the Hubble parameter~\cite{Niko1}
\begin{equation}
H(\phi)=\frac{c}{f'(\phi)}\sqrt{\frac{\phi f'(\phi)-3f(\phi)}{6}}
\label{Hub1}
\end{equation}
and the deceleration parameter~\cite{Niko2}
\begin{equation}
q(\phi)=\frac{2\phi f'(\phi)-3f(\phi)}{\phi f'(\phi)-3f(\phi)}.
\label{dec1}
\end{equation}
We also have the expression for the time dependence of $\phi$:~\cite{Niko1}
\begin{equation}
\dot{\phi}=\frac{\sqrt{6}c(\phi f'-2f)\sqrt{\phi f'-3f}}{2f'^2+\phi 
f'f''-6ff''}.
\label{phidot2}
\end{equation}
Combining Eqs.~(\ref{Hub1}--\ref{phidot2}) and using
$\dot{q}=\dot{\phi}q'(\phi)$ leads to
\begin{equation}
\frac{\dot{q}}{H}=\frac{18f'(\phi f'-2f)(\phi f'^2-\phi ff''-ff')}{(\phi f'-3f)^2(2f'^2+\phi f'f''-6ff'')}.
\label{dec2}
\end{equation}
From Eq.~(\ref{jer2}) we finally obtain
\begin{eqnarray}
& & j(\phi)=[2\phi^2 f'^4+10\phi^3 f'^3 f''-75\phi^2 f'^2 ff''-12\phi ff'^3+18f^2 f'^2 \nonumber \\
& & +189\phi f^2 f'f'' - 162f^3 f'']\times[(\phi f'-3f)^2(2f'^2+\phi f'f''-6ff'')]^{-1}.
\label{jer3}
\end{eqnarray}

We now examine the case $f(R)=R-\frac{\alpha^2}{3R}$, where $\alpha$ is
a constant, which is a possible
explanation of current cosmic acceleration~\cite{acc}.
In this model the present value of $\phi$ is $\phi_0=(-1.05\pm0.01)\alpha$,
where $\alpha=(7.35^{+1.12}_{-1.17})\times10^{-52}m^{-2}$\cite{Niko2}.
We do not need to know the exact value of $\alpha$ since it does not affect
non-dimensional cosmological parameters.
Substituting $\phi_0$ into~(\ref{jer3}) gives
\begin{equation}
j_0=1.01^{+0.08}_{-0.21}.
\label{jer4}
\end{equation}
This value does not overlap with the value 
$j=2.16^{+0.81}_{-0.75}$, obtained from the combination of three
kinematical data sets: the gold sample of type Ia supernovae~\cite{Gold},
the SNIa data from the SNLS project~\cite{SNLS}, and the X-ray galaxy cluster
distance measurements~\cite{jerk2}.
The origin of this disagreement could come from the assumption of
constant jerk used there.
However, two of the three data sets separately are consistent with the
$f(R)=R-\frac{\alpha^2}{3R}$ model: the SNLS SNIa set gives
$j=1.32^{+1.37}_{-1.21}$ and the cluster set gives
$j=0.51^{+2.55}_{-2.00}$, and it is the gold sample data
that yields larger $j=2.75^{+1.22}_{-1.10}$~\cite{jerk2}.~\footnote[2]{
The value $q_0=-0.81\pm0.14$ found in~\cite{jerk2}
from the combined three data sets agrees with $q_0=-0.67^{+0.06}_{-0.03}$
derived in the $f(R)=R-\frac{\alpha^2}{3R}$ model~\cite{Niko2}.
Each set separately agrees with our model as well.}

In the $f(R)=R-\frac{\alpha^2}{3R}$ model the deceleration-to-acceleration
transition occurred at $\phi_t=-\sqrt{5/3}\alpha$~\cite{Niko2}.
The cosmic jerk parameter at this moment can be found from
Eq.~(\ref{jer3}):
\begin{equation}
j_t=\frac{10}{9}.
\label{jer5}
\end{equation}
This value shows that the jerk parameter in $f(R)$ gravity changes
significantly between the deceleration-to-acceleration transition and now,
indicating the departure of $f(R)$ gravity models from
$\Lambda CDM$.
It would be interesting to generalize the kinematical approach
of~\cite{jerk2} to time dependent jerk and compare the results with $f(R)$
gravity models. 
More constraints on these models could also be provided by non-dimensional
parameters containing higher derivatives of the scale factor, such as the
snap parameter $s=\frac{\ddot{\ddot{a}}}{aH^4}$~\cite{snap}.

\section{Summary}

We derived the expression for the cosmic jerk parameter in
$f(R)$ gravity formulated in the Einstein gauge.
We used the Palatini variational principle to obtain the field equations
and apply them to a flat, homogeneous, and isotropic universe filled with dust.
The value of the jerk parameter for the particular case
$f(R)=R-\frac{\alpha^2}{3R}$ does not overlap with the value obtained
from cosmological data of the SNIa gold sample, but
is consistent with the values obtained from more recent SNLS SNIa data and
the X-ray galaxy cluster data~\cite{jerk2}.
Therefore, Palatini $f(R)$ models in the Einstein gauge,
including the case $f(R)=R-\frac{\alpha^2}{3R}$, provide a possible
explanation of current cosmic acceleration.
Further observations should give stronger constraints on $j$ and 
on $f(R)$ gravity.

\end{document}